\documentclass[conference]{IEEEtran}
\IEEEoverridecommandlockouts
\usepackage{cite}
\usepackage[numbers]{natbib}
\usepackage{amsmath,amssymb,amsfonts}
\usepackage{algorithmic}
\usepackage{graphicx}
\usepackage{textcomp}
\usepackage{xcolor}
\usepackage[utf8]{inputenc}

\usepackage{acronym}
\usepackage{url}
\def\BibTeX{{\rm B\kern-.05em{\sc i\kern-.025em b}\kern-.08em
		T\kern-.1667em\lower.7ex\hbox{E}\kern-.125emX}}
\begin{document}

\title{Dynamic Social Interaction Mechanics in CrossAnt\\
}

\author{
	\IEEEauthorblockN{Samuel Gomes}
	\IEEEauthorblockA{
	    \textit{Intelligent Agents and}\\ \textit{Synthetic Characters Group}\\
		\textit{INESC-ID \& Instituto Superior}\\ \textit{T\'{e}cnico, Universidade de Lisboa}\\
		Lisbon, Portugal\\
		samuel.gomes@tecnico.ulisboa.pt
	}
	\and
	\IEEEauthorblockN{Carlos Martinho}
	\IEEEauthorblockA{
	    \textit{Intelligent Agents and}\\ \textit{Synthetic Characters Group}\\
		\textit{INESC-ID \& Instituto Superior}\\ \textit{T\'{e}cnico, Universidade de Lisboa}\\
		Lisbon, Portugal\\
		carlos.martinho@tecnico.ulisboa.pt
	}
	\and
	\IEEEauthorblockN{João Dias}
	\IEEEauthorblockA{
	    \textit{Intelligent Agents and}\\ \textit{Synthetic Characters Group}\\
		\textit{INESC-ID \& Instituto Superior}\\ \textit{T\'{e}cnico, Universidade de Lisboa}\\
		Lisbon, Portugal\\
		joao.dias@tecnico.ulisboa.pt
	}
}

\maketitle

\begin{abstract}
	Nowadays, big effort is being put to study gamification and how game elements can be used to engage players. In this scope, we believe there is a growing need to explore the impact game mechanics have on the players' interactions and perception.
	This work focuses on the application of game mechanics to lead players to achieve certain types of social interaction (we named this type of mechanics \textit{social interaction mechanics}). 
	A word matching game called \textit{CrossAnt} was modified so that it could dynamically generate different \textit{social interaction mechanics}. These mechanics consisted in different key combinations needed to play the game and were aimed to promote what we think are three important types of social interactions: \textit{cooperation}, \textit{competition} and \textit{individual exploration}.
	
	Our evaluation consisted on the execution of several sessions where two players interacted with the game for several levels and had to find for themselves how to perform the actions needed to succeed. While some of the levels required the input from both players in order to be completed, others could be completed by each player independently.
	Our results show that cooperation was perceived when both players had to intervene to perform the game actions. However, longer interactions may still be needed so that the other types of interactions are promoted.
\end{abstract}


\maketitle


\acrodef{LM-GM}[LM-GM]{Learning Mechanics - Game Mechanics}
\acrodef{GOM}[GOM]{Game Object Model}

\acrodef{ITS}[ITS]{Intelligent Tutoring System}
\acrodef{ATES}[ATES]{Advanced Topics in Entretainment Systems}

\acrodef{MOJO}[MOJO]{Montra de Jogos}

\section{Introduction}

Big importance has been given to define and analyze the way people experience games \cite{deterding2011game}.
In fact, as social interactions help to define experiences, several studies approached how certain behaviors can be promoted in groups \cite{consalvo2011using, oksanen2014game}. There is however much to explore in this field, and for that reason a new direction of research is identified: \textit{the generation of certain types of mechanics at certain moments in order to lead players to achieve specific types of social interactions}.
We support this idea by categorizing the mechanics which aims to promote a certain kind of social interaction among the players as \textit{social interaction mechanics}. This definition is divergent from the existing ones which mainly focused on the synergy between the players and the game \cite{sicart2008defining, martinho14}.

Acknowledging this idea, we aim to approach the following general problem:
\\
\\
\textit{How can the generation of certain types of mechanics lead players to achieve specific types of social interactions?}
\\

To approach this problem, we modified a word matching game called \textit{CrossAnt} so that the keyboard bindings needed to play the game were dynamically generated. These keyboard bindings could either target the players individually or in conjunction. The dynamic generation of key bindings can be considered an interaction mechanic, acknowledging that our hypothesis are valid:
\\
\begin{itemize}
    \item
    H1: Using key mappings which require input from only one player (single-player mappings), individual exploration or competition can be promoted;
    \item
    H2: Using mappings which require input from both players (multi-player mappings), cooperation can be promoted;
    \item
    H3: Different social interactions types might not be promoted in the same moments of the interaction;
    \item
    H4: Social interactions may not be perceived in early gameplay stages.
\end{itemize}

The rest of the document is outlined as follows: first, the related work is presented, namely some studies regarding the promotion of certain types of interactions in games; Then, in "Work Description", the game \textit{CrossAnt} as well as the changes made to it are detailed; Finally, the results of the conducted tests are provided and discussed in sections "Work Evaluation" and "Discussion".

\section{Related Work}

\subsection{The importance of playfulness}

Recent research has been conducted to define ``gamification" \cite{deterding2011game}, and to distinguish gamefulness and from playfulness.
\textit{playing} and \textit{gaming} are referred as divergent concepts, in the sense that the word play is more general and extends to much more than an isolated action, is actually a state of mind.
Oppositely to what is commonly thought, the act of ``playing a game" can be considered to continue after the interaction ends, for example if we dream about some of its aspects. Moreover, we can argue that when we watch someone else play, we are also ``playing the game" through the actions of the other player. What we want to point out is that \textit{what really matters is the whole experience.}
As social interactions are crucial to define experiences, methods for promoting them in certain moments of the gameplay are very important to improve the design of interactive applications such as games.

\subsection{Studying Social Interactions in Games}
Some work explored how nowadays games influence certain types of behaviours in the players. 
On the one hand, \citeauthor{consalvo2011using} checks how Facebook and other social network oriented games promote certain types of interaction \cite{consalvo2011using}. \citeauthor{consalvo2011using} arguments the displaying of a friends list to promote interaction and awareness between players, using rewards for visiting others to encourage players to explore their friends’ spaces or leaderboards to promote competition. 
On the other hand, Mechanics tailored to collaboration were proposed by \citeauthor{oksanen2014game} \cite{oksanen2014game} like the use of shared space and objects between the players, complementary actions (actions which can only be made by a group) and indirect actions (situations where some of the players have a task that requires other player's action).
This work aims to address the promotion of specific behaviors by applying social interaction mechanics in different moments. 

\subsection{Characterizing Interactions}

The related work explored mechanics to promote either collaboration, cooperation  or competition. While cooperation and collaboration involved allowing the players to have dependent actions or be rewarded when executing a common goal, competition involved allowing the players to have a common goal but only rewarding one of them, like in tournaments. These dimensions of interaction are also identified by work approaching the dynamics of multiplayer serious games \cite{Wendel2016}.

As the game we used implies the concurrent execution of actions between the players, from the presented interaction types, we chose to consider \textit{cooperation} and \textit{competition}. Besides, we acknowledge that in certain moments of the interaction, a player interacting with our game could feel that the intervention from the other player is not necessary for his/her own success and that he/she feels that both players are performing concurrent individual tasks. As such, we added \textit{individual exploration} to the two above mentioned types of interaction.



\section{Work Description}

\begin{figure}
	\centering
	\includegraphics[width=0.9\linewidth]{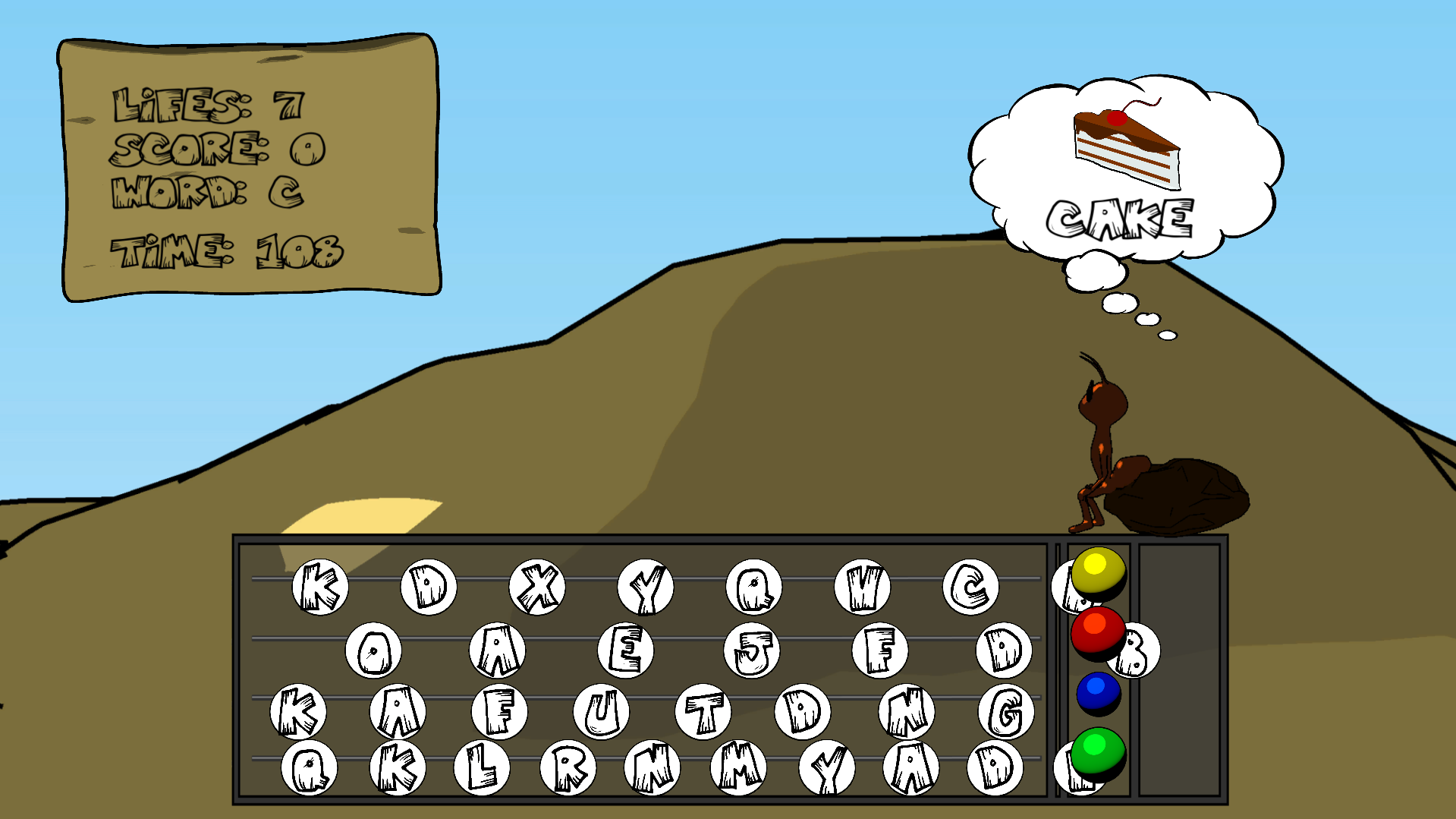}
	\caption{\label{fig:crossAntScreenshot}Screenshot of the original version of \textit{Cross Ant}. In the top left panel, the player can check several game state metrics like his/her remaining lives, the current score and the time left to complete the level. The Queen ant is included on top right of the track along with her request in a balloon. The letters, which move from left to right above the track have to be selected by triggering the coloured stamps on the right. In the screenshot, the blue stamp is triggered.}
\end{figure}

Considering the presented problem and hypothesis, we focused this work in the modification of a word matching game called \textit{CrossAnt} initially developed during the marathon \textit{Global Game Jam 2018}\footnote{The original version of the game can be found at \url{https://globalgamejam.org/2018/games/crossant} (verified in 5 of November of 2018) and the current version can be played at \url{https://samgomes.github.io/interaction-mechanics-cross-ant/} (verified in 5 of November of 2018)}.The original version of the game is presented in Fig. \ref{fig:crossAntScreenshot}.
In this game, the player assumes a role of helper ant which has to serve a \textit{Queen Ant} the food she requests in a limited amount of time. In order to get some food, the player has to use colored stamps positioned at the bottom right side of the screen to select the letters of each request in a rolling letter soup. If the player selects a letter which is not the next one to complete the request, he/she looses a life. When a certain amount of lives are lost, the game ends. While the game is not over, it continues to generate new words for the player to complete, increasing the speed at which the letters move in each level.

\begin{figure}
	\centering
	\includegraphics[width=1.0\linewidth]{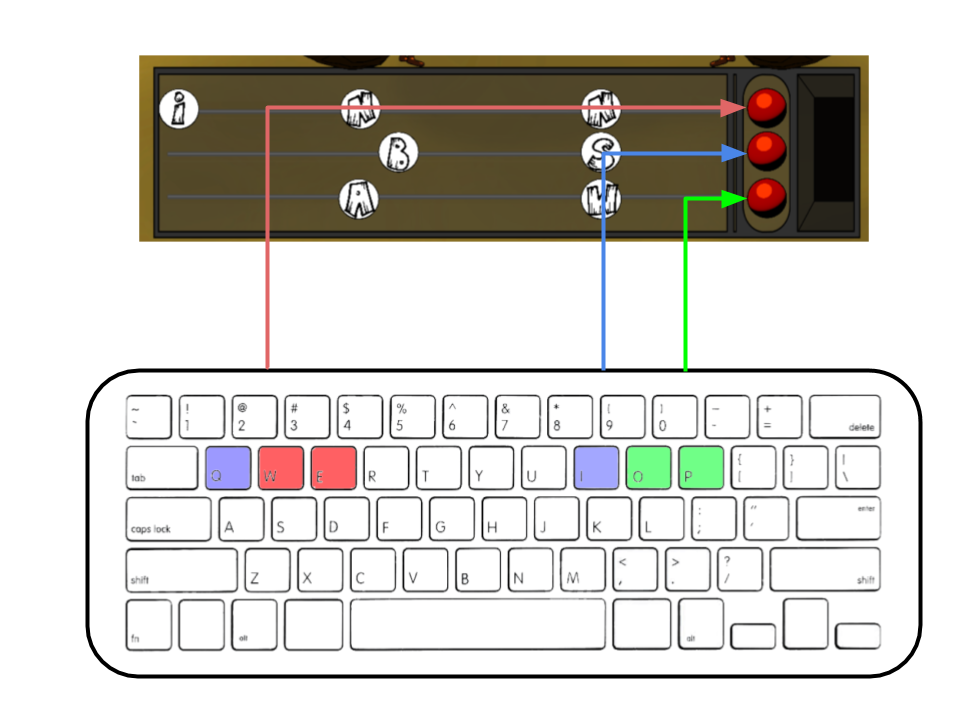}
	\caption{\label{fig:combinationMapping}Possible mapping types (represented as different colors) between controls keys (in this case the keyboard is the controller) and game stamps. \{W,E\} and \{O,P\} are examples of single-player key combinations which are connected to the top and bottom game stamps. The multi-player key combination \{Q,I\} maps to the middle stamp.}
\end{figure}

\begin{figure}
	\centering
	\includegraphics[width=0.8\linewidth]{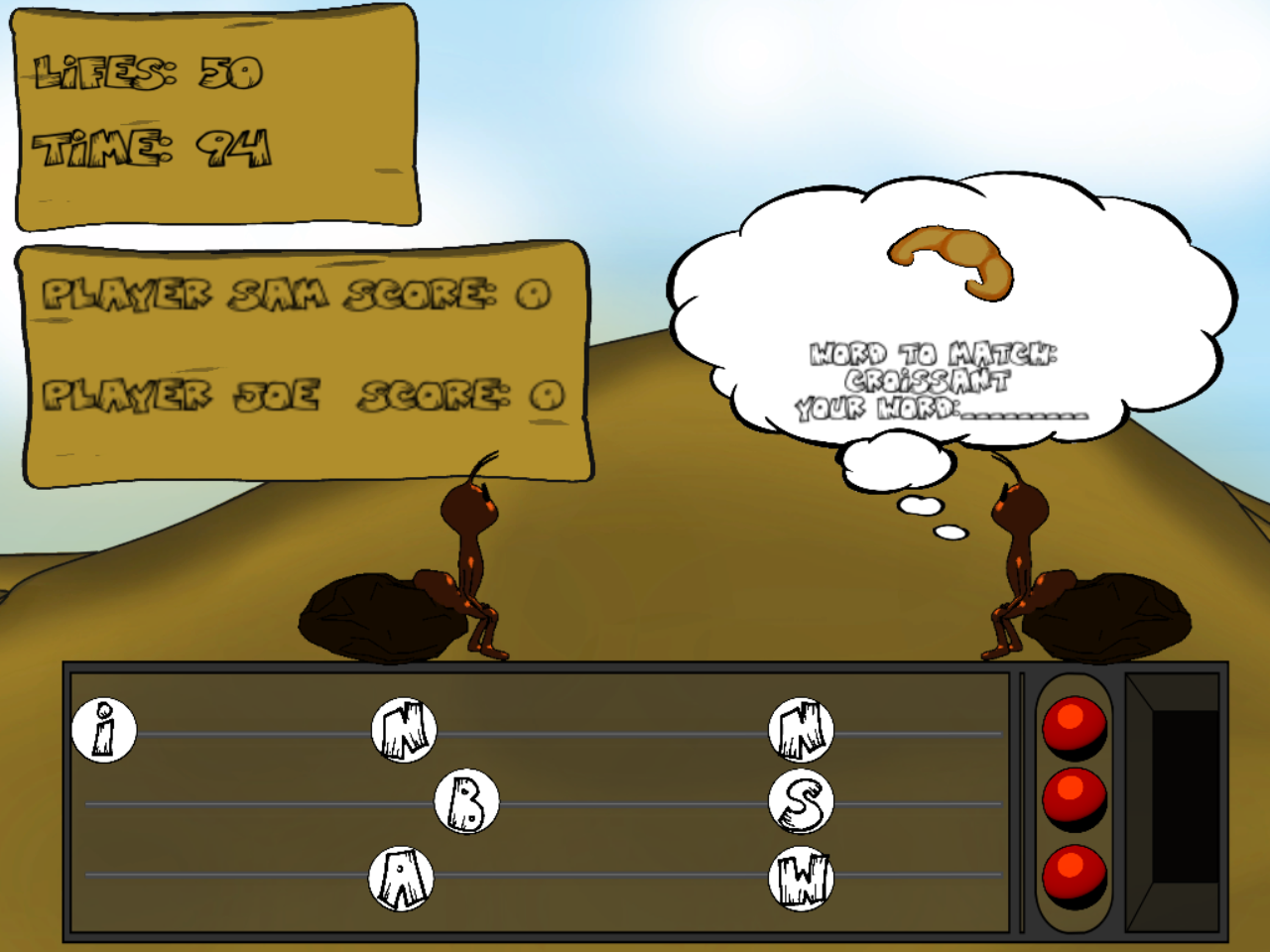}
	\caption{\label{fig:modifiedCrossAntScreenshot}Screenshot of the modified version of \textit{Cross Ant}. The game textures were slightly changed. Individual score panels and an additional ``Queen Ant" were also included. The word to match was moved to the top of the current partial word on the balloon of the left ``Queen Ant". The track was reduced to include only three letter lines and stamps. All of the stamps were painted in red.}
\end{figure}

After improving some visual aspects which were not fully developed in the original version 
and coloring all stamps equally to not misguide players\footnote{Colors were only relevant when using an xbox controller in the original version}, some changes were made so that the our research problem was considered. The modified version of the game is included in Fig. \ref{fig:modifiedCrossAntScreenshot}. The functionalities added to the modified version include:
\begin{itemize}
	
	\item
	Allowing two players to interact concurrently (local multiplayer); Acknowledging this configuration, two individual scores were added to the score panel and two ``Queen ants" were added to the center of the screen. The last change was included to allow two helper ants to move in opposite directions and deliver food without overlapping;
	
	
	\item 
	Reducing the number of the game stamps to three in order to simplify the layout of the game;
	
	\item
	Adding several types of mappings between the keys and the game buttons, dynamically generating them in each level. Two mapping types were considered: \textbf{single-player key mappings} (each player can find the key combinations for the game buttons using his/her keys) and \textbf{multi-player key mappings} (the combination requires input from the two players). Examples of the key combinations produced by each of these mappings are depicted in Fig. \ref{fig:combinationMapping}. This was the most important change as it was directly aimed to approach the initial hypothesis.

\end{itemize} 

Through the inclusion of individual scores and two separate ``Queen Ants", we could reinforce the players to perceive the presence of competition or that concurrent individual actions were being performed on the levels where single key mappings were generated.

\section{Work Evaluation}

Firstly, several pilot tests were conducted in order to check the impact our modifications had in the overall playability of the game.
In these tests, 3 groups were asked to play alongside the researchers and to identify possible negative aspects associated with the gameplay. The groups were also asked to provide some hints which could be used to improve or fix the identified problems. 
Based on the provided feedback, some further fixes were made such as a reduction of the track speed and a method to guarantee that valid key combinations could be found for all levels.

Afterwards, a set of tests was conducted where several groups of two elements played the game at the event \ac{MOJO}\footnote{\url{https://tecnico.ulisboa.pt/pt/tag/mojo/} (verified in 5 of November of 2018)} held at the college Instituto Superior Técnico. This event allows college students to present their games to other students and outside visitors.  
A description of the evaluation methodology, as well as the results extracted from the tests are included in the next section.

\begin{figure}
	\centering
	\includegraphics[width=0.8\linewidth]{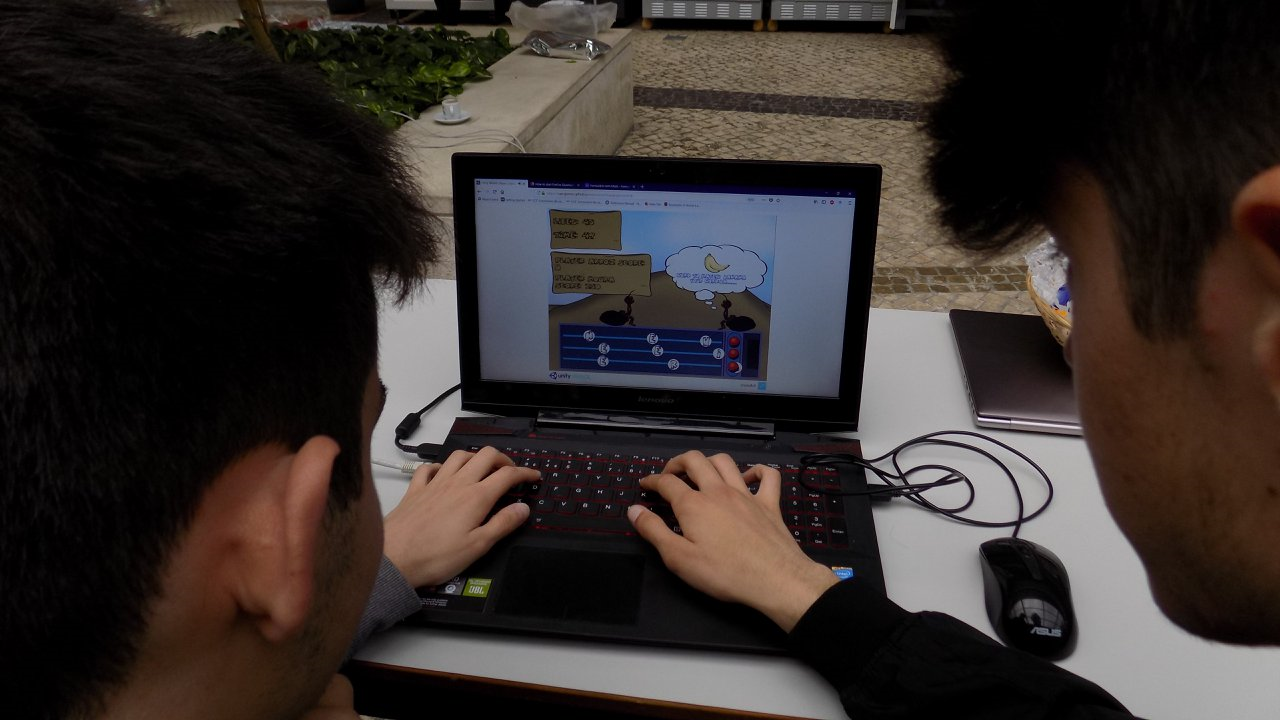}
	\caption{\label{fig:experimentBeingRun}Photo of a group playing the modified \textit{CrossAnt} at \ac{MOJO}.}
\end{figure}

After playing the game for several levels, each group was asked about their experiences through questionnaires. Namely, each participant was asked to rate his/her experience while playing each game level in a Likert scale, ranging from 1 to 5, regarding three social interactions: cooperation (the participant felt that the members helped each other throughout the level), competition (the participant felt that he/she competed with the other participant throughout the level) and individual exploration (the participant felt that the other participant's actions were not relevant for the completion of the level). A photo of a group playing the game at \ac{MOJO} is included in Fig. \ref{fig:experimentBeingRun}.

Although mostly good feedback was provided throughout the experiments, some critiques were received while conducting the extensive tests at \ac{MOJO} focusing on the game having a big learning curve due to the varying mechanics.

\subsection{Results}

After filtering invalid questionnaires data, responses from 33 people were recovered. More specifically, data from 203 played levels was considered.
Statistical tests were conducted based on the scores participants gave on competition, collaboration/collaboration and individual experience while playing each of the game levels. 

Remembering our hypothesis, the objectives of the statistical tests were to check if significant score differences could be observed when considering different key mapping types (different interaction mechanics) and different moments of the interaction.

\subsection{Checking differences between types of key mappings}

First, we compared the scores given by each participant when facing one of our mapping types.
We considered not only the first and last played levels, but also the means of the scores along the interaction.
The distributions were checked to be non parametric and for that reason, Wilcoxon Signed-ranks tests were performed. The results of executing such tests are presented in the tables \ref{table:wilcoxonTestsFirstLevels}, \ref{table:wilcoxonTestsLastLevels} and \ref{table:wilcoxonTestsMeanScores}.

The data provided by the tables indicates that participants perceived the presence of cooperation while playing the last levels using multi-player mappings ($Z= 1,994, p<0.05, r= 0.425$). This tendency was even more significant when grouping the data using the means of the scores throughout the levels ($Z= 2.68, p< 0.01, r= 0.571$). 
Oppositely, no significant differences are shown between key mapping strategies on the first recorded levels.
These results support \textit{H2}, but not \textit{H1}.

\begin{table}[]
	\caption{Wilcoxon Tests (Multi vs Single player mappings) - First Played Levels}
	\label{table:wilcoxonTestsFirstLevels}
	\begin{tabular}{llll}
		& Collaboration & Competition & Indiv. Exp. \\
		Z & -0,969 & -1,600 & -1,245\\
		Significance Sig. (bilateral) & 0,333 & 0,110 & 0,213 \\
		r&-&-&-
	\end{tabular}
\end{table}
		
\begin{table}[]
	\caption{Wilcoxon Tests (Multi vs Single player mappings) - Last Played Levels}
	\label{table:wilcoxonTestsLastLevels}
	\begin{tabular}{llll}
		& Collaboration & Competition & Indiv. Exp.  \\
		Z & -1,994 & -1,329 & -1,098\\
		Significance Sig. (bilateral) & 0,046 & 0,184 & 0,272 \\
		r&0.425&-&-
	\end{tabular}
\end{table}

\begin{table}[]
	\caption{Wilcoxon Tests (Multi vs Single player mappings) - Mean Scores of all Played Levels}
	\label{table:wilcoxonTestsMeanScores}
	\begin{tabular}{llll}
		& Collaboration & Competition & Indiv. Exp.  \\
		Z & -2,680 & -0,782 & -1,728\\
		Significance Sig. (bilateral) & 0,007 & 0,434 & 0,084\\
		r&0.571&-&-
	\end{tabular}
\end{table}

\subsection{Checking differences between moments of the interaction}
A second test was executed in order to check the differences between scores in the first and last played levels. As before, the distributions were checked to be non-parametric and so Wilcoxon Signed-ranks tests were considered. The results are presented in the table \ref{table:wilCoxonTestsEvolution}. We can observe that all differences are not statistically relevant. This indicates that no bias to any interaction was seen over both strategies. However, if we only considering multi-player key mappings (table \ref{table:wilCoxonTestsEvolutionMixedKeys}), we can observe that collaboration scores changed throughout the course of the game. It was this change which allowed the significant difference between key mapping types only in the last interactions presented on the previous test, and suggests that the scores changed midway through the interactions.
This conclusion supports \textit{H3} and \textit{H4}.
	

\begin{table}[]
	\caption{Wilcoxon Tests (First vs Last played levels)}
	\label{table:wilCoxonTestsEvolution}
	\begin{tabular}{llll}
		& Collaboration & Competition & Indiv. Exp.  \\
		Z & -1.664 & -0.312 & -0.943\\
		Significance Sig. (bilateral) & 0.096 & 0.755 & 0.350\\
		r&-&-&-
	\end{tabular}
\end{table}

\begin{table}[]
	\caption{Wilcoxon Tests (First vs Last played levels) - Multi-player mappings}
	\label{table:wilCoxonTestsEvolutionMixedKeys}
	\begin{tabular}{llll}
		& Collaboration & Competition & Indiv. Exp.  \\
		Z & -2,9 & -0,320 & -0.439\\
		Significance Sig. (bilateral) & 0,004 & 0,749 & 0,661\\
		r&0.569&-&-
	\end{tabular}
\end{table}

\section{Discussion}

By analysing the results, we can perceive that although cooperation being successfully incentivized, the same was not seen on the other types of behavior like competition or individual experience. Such results are possibly due to the common goals of each game level. Remembering that players have to cooperate in order to complete each word and survive through the game, maybe this is the predominant type of social behavior while playing it and so is the first to be incentivized. However, significant differences were not perceived right away. In fact, after contrasting the scores given in the first and last played levels for multi-player mappings, it can be concluded that some practice was needed before cooperation was perceived. Maybe one tutorial level was not enough for the players to adapt to the game. Indeed, people who experienced the game in \ac{MOJO} referred that it had a big learning curve due to the dynamically generated key bindings.

As such, we argue that with longer playing times, other behaviors like individual experience or competition can also be promoted using the single player bindings. It is possible that these behaviors are initially overwhelmed by the importance given to the common goals when people engage in mixed types of interactions in a game such as the one tested. In the modified \textit{CrossAnt}, both players try to complete the word in order to advance in the game, and so they tend tend to exploit the controls as a group before trying other strategies or checking how high their score is. With few played levels, there is no need to compete or to try to solve each level alone because both players are still trying to get the word right before the time ends.

\section{Conclusion}
In this work we provided a simple and creative way to study the interactions occurring between a game and its players and evaluated if such process helped to promote certain social behaviors between the players in certain moments of the interaction. 
Some modifications were made to a word matching game called \textit{CrossAnt} and some tests were conducted in order to measure the impact such modifications had on the players' social interactions.
The results revealed that collaboration was promoted when input was required from both players, although this only emerged mid way through the interactions, which reveals that the promotion of certain kinds of interaction might, in some cases like ours, require some practice time in order for the players to get used to the game.

In the future, it is possible to continue this study, checking what can be verified when measuring longer interactions and using other mechanics, for example transmitting in-game feedback such as warnings or complements. 

Another way to extend this study is to apply the same dynamic key mapping approach to games which provide long interactions in nature, like open world games, real time strategy games or role playing games.

\bibliographystyle{IEEEtranN}
\bibliography{IEEEabrv,References}

\end{document}